# Generation of Biometric Key for Use in DES

Rupam Kumar Sharma
Don Bosco College Of
Engineering and Technology,
Assam, India
sun1_rupam1@yahoo.com

## ABSTRACT
Cryptography is an important field in the area of data encryption. There are different cryptographic techniques available varying from the simplest to complex. One of the complex symmetric key cryptography techniques is using Data Encryption Standard Algorithm. This paper explores a unique approach to generation of key using fingerprint. The generated key is used as an input key to the DES Algorithm

## Keywords
Biometrics, Fingerprint, Encryption, Symmetric, Minutiae - points.

## 1. INTRODUCTION
Biometrics is the science of identification of humans by their characteristics. Fingerprints are the most widely used parameter for human identification amongst all biometrics. Fingerprint comprises a set of minutiae. The orientation of the minutiae uniquely identifies a fingerprint. This feature has been exploited to generate a unique code for an individual. The code is further modified to obtain a 56 bit key for use in the symmetric key encryption namely Data Encryption Standard(DES). DES is one of the mostly used symmetric key cryptography algorithm in the present day apart from the Advanced Encryption Standard(AES).This encryption technique can be used to encrypt a block of characters. A ridge in fingerprint is defined as a single curve segment and a valley is the region between two adjacent ridges. The collective set of ridge endings and bifurcations form the minutiae. The minutiae can be of different types including dots ,islands, ponds or lake, spurs ,bridges and crossovers.

## 2. FINGERPRINT PROCESSING
### 2.1 Histogram Equalization
Histogram equalization[1] increases the contrast of images. In this technique the basic idea here is to map the gray levels based on the probability distribution of the input gray levels. Histogram Equalization transforms the intensity values of the image as given by equation

$S_k = T(r_k) = \sum P_r(r_j) = \sum n_j/n$ for j=1..k.

Where $S_k$ is the intensity value in the processed image corresponding to intensity $r_k$ in the input image and $p_r(r_j)$=1,2,3…L is the input fingerprint image intensity level.

### 2.2 Binarization
Fingerprint image Binarization [2] is to transform the 8-bit Gray fingerprint image with 0-value for ridges and 1 value for furrows. After the operation, ridges in the fingerprint are highlighted with black color while furrows are white.

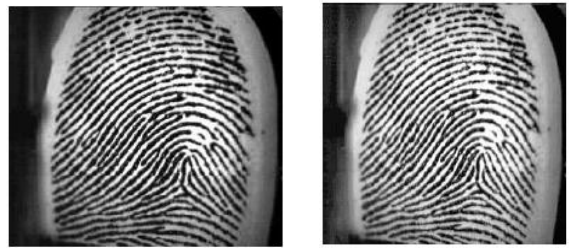

**Fig1. sample fingerprint image    Fig 2.After Histogram Equalization**

### 2.3 Morphological Operation
Morphological Operation[4]are used to understand the structure or form of an image. It means identifying objects or boundaries within an image. There exists three primary morphological functions-erosion, dilation, hit or miss. Morphological operations are performed on binary images where the pixel values are either 0 or 1.Binary morphological operators are applied on binarized fingerprint image to remove spurs, bridges ,line breaks etc. A process called thinning is also applied to reduce thickness of lines. It is a process particularly used for skeletonisation.

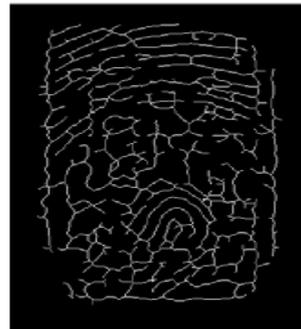

**Fig3: Morphological Operation**

In this mode it is commonly used to tidy up the output of edge detectors by reducing all lines to single pixel thickness. The output of thinning produces another binary image.

### 2.4 Minutiae points' extraction

The binary image is thinned such that a ridge is only one pixel wide. Among all fingerprint features ,minutia point features with corresponding orientation maps are unique enough to discriminate amongst fingerprint robustly; the minutiae feature representation reduces the complex fingerprint recognition problem to a point pattern matching problem. The minutiae are extracted from the enhanced, thinned and binary image. One of the minutia extraction technique is crossing number.

## 2.4.1 Crossing Number

This method involves the use of skeleton image where the ridge flow pattern is eight-connected. The local neighborhood of each ridge pixel in the image is scanned out using a 3x3 window.

**Table 1. A 3x3 neighbourhood**

| $P_4$ | $P_3$ | $P_2$ |
|---|---|---|
| $P_5$ | P | $P_1$ |
| $P_6$ | $P_7$ | $P_8$ |

The crossing number(CN) value is then computed as follows

$CN = 0.5 \sum |P_I - P_{I+1}|$ for i=1... 8

Where $P_9 = P_1$. It is defined as the half the sum of the differences between pairs of adjacent pixels in the eight-neighborhood. Using the properties of CN as mentioned below, ridge pixel can be classified as ridge ending, bifurcation or non-minutiae point.

**Table 2. Properties of Crossing Number**

| CN | Property |
|---|---|
| 0 | Isolated point |
| 1 | Ridge ending point |
| 2 | Continuing ridge point |
| 3 | Bifurcation point |
| 4 | Crossing point |

A pixel is thus classified as a ridge ending if it has only one neighbouring ridge pixel in the window, and classified as a bifurcation if it has three neighboring pixel and continuing ridge point if it has two neighboring pixels etc.[5]

## 3 Data Encryption Standard

The Data Encryption Standard(DES) is a symmetric-key block cipher published by the National Institute of Standards and Technology.(NIST).The encryption process is made of two permutations(p-boxes),which we call initial and final permutations and sixteen Feistel rounds. Each round uses a different 48-bit round key generated from the cipher key according to a predefined algorithm.Des uses 16 rounds.Each round of DES is a Fiestel Cipher.The heart of DES is the DES function.The DES function applies a 48-bit key to the rightmost 32bits to produce a 32bit output.The function is made up of four sections -and expansion P-Box, a whitener, a group of S-boxes and a straight P-Box.

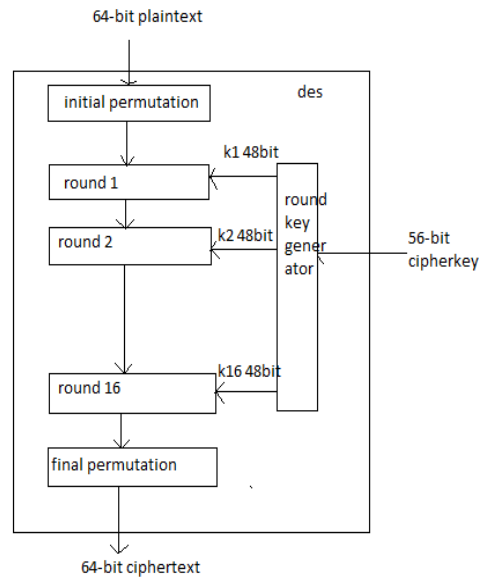

**Fig4: General Structure of DES.**

## 4 Proposed Model

The proposed Model takes as input the given JPEG/JPG image and gives as output a 64-bit key. The key generated is input to the parity drop table DES key generator. The entire focus of the paper is the second block i.e 64-bit key generator from the

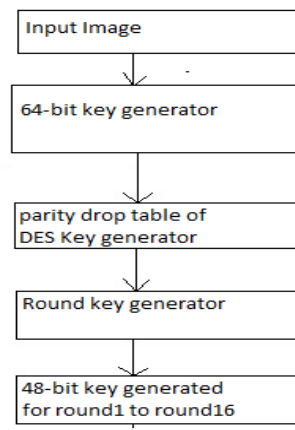

**Fig5: Block diagram of the proposed model**

Input image. The 64-bit key generator block is further divided into sub-blocks portraying in details the inner working of the block. The input JPEG/JPG is converted to binary image. The binary images with only two levels of interest.The black pixels that denote ridges and the white pixels that denote valleys are employed by almost all minutiae extraction algorithms. A grey level image is translated into a binary image in the process of binarization, by which the contrast between the ridges and valleys in a fingerprint image is improved. Thinning process is performed to reduce thickness of lines. Thinning is a morphological operation that is used to remove selected foreground pixels from binary images. It is particularly used for skeletonisation. It is used to tidy up the output of edge detectors by reducing all lines to single pixel thickness. [4]

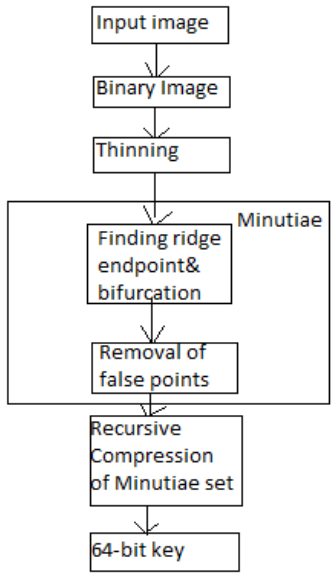

**Fig6: Details of 64-bit key generator**

After the fingerprint ridge thinning marking minutia points is relatively easy. For each 3X3 window ,if the central pixel is 1 and has exactly 3 one-value neighbors, then the central pixel is a ridge branch. If the central pixel is 1 and has only 1 One-value neighbor, then the central pixel is a ridge ending. The false ridge breaks due to insufficient amount of ink and ridge cross-connections due to over inking are not totally eliminated. These false minutiae might impact the accuracy and genuine of the finger code generated from a given fingerprint image. These false minutiae are removed. The set of minutiae set generated is of greater size. This set need to be reduced to derive a 64-bit set. The original minutiae set is recursively reduced until a 64-bit key is obtained.

### 4.1 Algorithm for generation of key

Step1:- NP ← No of minutiae points in the region of interest.
Step2:- Rem ← NP mod 64
Step 3:- NP ← NP – Rem //this makes NP perfect divisible by 64.
Step 4:- I ← NP /64//gives number of recursive iteration to perform compression of the key size to 64-bit.
Step 5:- for 1 to I
    do
Drop Left 32 bit and Right 32 bit.
Divide the remaining key set into $K_L$ and $K_R$
Swap $K_L$ and $K_R$.
done.
End of for.

## Practical Implementation and Test Results

The above proposed algorithm has been implemented and tested successfully in Matlab. Following shows the snapshot of the implementation.

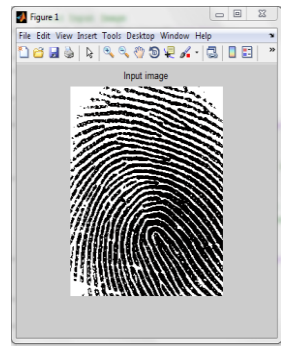 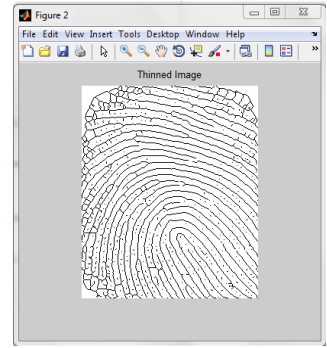

**Fig6:- Input Image**         **Fig7:- Thinned Image**

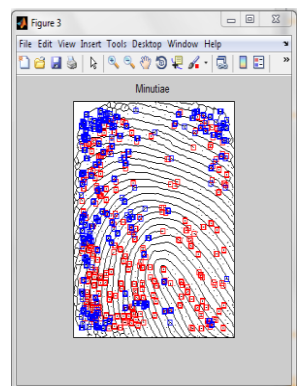

**Fig8:- Minutiae Set of the Input Image identified by blue and red mark.**

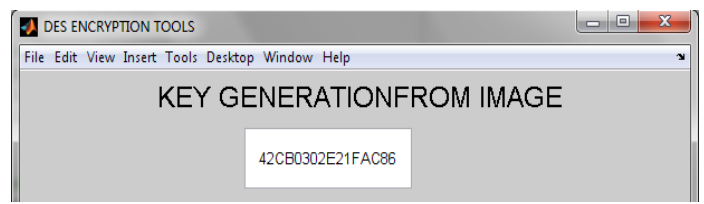

**Fig9:-16digit Hexadecimal key generated from Input Image using algorithm of 4.1. This key is input to DES.**

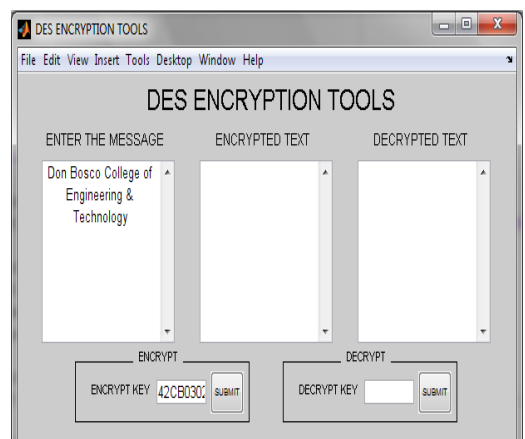

**Fig10:- GUI Interface to input text, key and display the corresponding encrypted text generated using DES.**

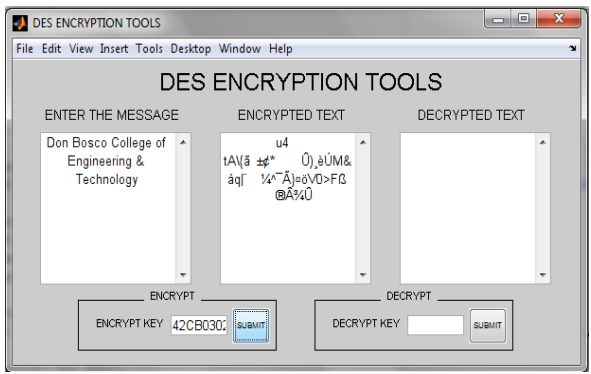

**Fig11:-Encrypted text generated with key 42CB0302E21FAC86**

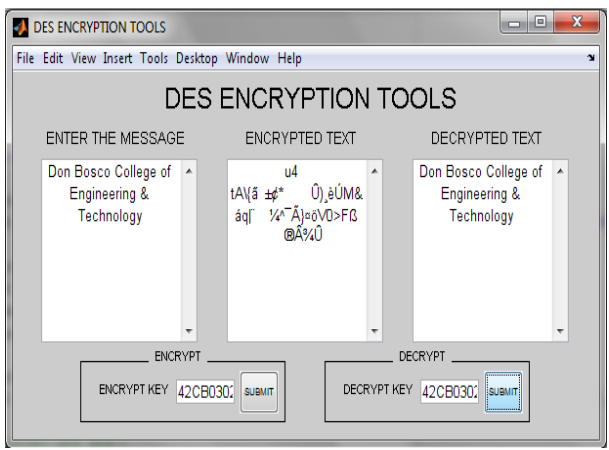

**Fig 12:- Decrypted text when symmetric key is supplied .**

## Conclusion:-

DES is one of the most widely used symmetric encryption algorithm. This paper describes an independent approach in generating the key from human fingerprint to be used in DES encryption. The algorithm is successfully implemented and tested in Matlab. The paper also describes some of the commonly used technique for generation and extraction of minutiae points from fingerprint.

## REFERENCE:-

[1] A Jagadeesan,Dr K.Duraiswamy "Secured Cryptographic key generation from multimodal Biometrics:Feature Level Fusion of Fingerprint and Iris" in International Journal of computer science and information security,Vol 7,No 2 ,February 2010.

[2] C Nandini and B.Shylaja "Efficient Cryptographic key generation from fingerprint using symmetric hash functions" in International Journal of Research and reviews in Computer Science.Vol 2,No 4 ,August 2011,ISSN:2079-2557.

[3] Ashwini R.Patil,Mukesh A Zaveri,"A Novel Approach for Fingerprint Matching using Minutiae",IEEE Fourth Asia International Conference on Mathematical/Analytical Modelling and computer Simulation,2010.

[4] R Seshadri,T.Raghu Trivedi,"High Secured Biometric Key generation system for data transfer",International Journal of Research and Reviews in Computer Science,Vol2,No1,March 2011.

[5] Roli Bansal,Priti Sehgal,PunamBedi in"Minutiae Extraction from fingerprint images- a Review" on International journal of computer science issues,vol 8,issue 5,no 3,September 2011.

[6] Federal Information Processing standards publication. U.S Department of commerce /National Institute of Standards and Technology.

[7] Cryptography & Network Security "Behrouz A.Forouzan.

[8] Naveena Marupudi,Eugene John and Fred Hudson in "Fingerprint Verification in Multimodal Biometrics"

[9] R.Sashank Singhvi,SP. Venkatachalam and others in "Cryptography Key Generation using Biometrics"

[10] Yusupov S. Yu, Medetov S.K. in "Application of Biometric Methods in Crytography"